\documentclass[12pt, preprint]{aastex}
\usepackage{emulateapj5}  

\def\ergcm2s{~erg cm$^{-2}$ s$^{-1}$ } 
\def\erg{~erg s$^{-1}$}		
\def\etal{et al.~}		

\def\msun{~M$_{\odot}$}

\def\n4038{~NGC4038/39}		

\def\x2{$\chi^{2}$}	



\begin{document}

\title{X-raying Chemical Evolution and Galaxy Formation in The Antennae 
 \\}

\author{ G. Fabbiano$^1$,  A. Baldi$^1$, A. R. King$^2$, T. J. Ponman$^3$,
J. Raymond$^1$, A. Read$^3$, A. Rots$^1$, Fran\c cois Schweizer$^4$, A. Zezas$^1$}
\affil{$^1$Harvard-Smithsonian Center for Astrophysics, 60 Garden
Street, Cambridge, MA 02138;  gfabbiano@cfa.harvard.edu; azezas@cfa.harvard.edu,
arots@cfa.harvard.edu}
\affil{$^2$Theoretical Astrophysics Group, University of Leicester, Leicester 
LE1 7RH, UK; ark@astro.le.ac.uk}
\affil{$^3$ School of Physics \& Astronomy, University of Birmingham, Birmingham 
B15 2TT, UK; tjp@star.sr.bham.ac.uk}
\affil{$^4$Carnegie Observatories, 813 Santa Barbara St., Pasadena, CA 91101-1292;
 schweizer@ociw.edu}

\shorttitle{X-raying The Antennae}
\shortauthors{Fabbiano et al.}
\bigskip

\begin{abstract}
We present the integrated 411~ks {\it Chandra} ACIS-S exposure of the Antennae
galaxies (NGC~4038/39). Besides a rich population of point-like sources, 
this spectacular image reveals a spatially and spectrally complex hot 
diffuse gaseous component. For the first time we detect intense line
emission from Fe, Ne, Mg and Si in The Antennae, and obtain a detailed picture 
of spatially varied metal abundances in the hot interstellar medium (ISM) of
a galaxy. In certain regions, the abundances of $\alpha$-elements may be many
times solar, while the Fe abundance is sub-solar or near-solar. The
differences in the local metal enrichment of the hot ISM may be related 
to the local star formation rates and to the degree of confinement of the
enriched hot ISM.
We also report large-scale gaseous features, including two gigantic,
$\sim$10-kpc-scale `loops' extending to the South of the merging disks,
and a low-surface-brightness hot halo, extending out to $\sim$18~kpc.
These features may be related to superwinds from the starburst in The Antennae
or result from the merger hydrodynamics. Their long cooling times suggest that
they may persist to form the hot X-ray halo of the emerging elliptical galaxy.

\end{abstract}

\section{Introduction}
The interacting pair of galaxies NGC~4038/39 (The Antennae), 
at a distance of 19~Mpc ($H_0 = 75$), has been studied intensely 
at all wavelengths as the nearest example of galaxies undergoing a 
major merger (e.g., Whitmore et al.\ 1999; Hibbard et al.\ 2001;
Neff \& Ulvestad 2000; Wilson et al.\ 2000; Fabbiano et al.\ 2001). 
The Antennae provide a local laboratory where astronomers
can easily observe phenomena that occur in the deeper Universe.  There,
merging is common and may be an important step in the evolution
of galaxies (e.g., Navarro, Frenk \& White 1995). 

The Antennae have been studied in X-rays since their first observation with {\it Einstein} 
(Fabbiano \& Trinchieri 1983). The first {\it Chandra} ACIS-S (Weisskopf et al.\
2000) observation of this system in December 1999
revealed both a population of exceptionally luminous point-like sources 
and complex hot diffuse emission with temperatures
in the range $kT\approx$ 0.3--0.8~keV.
This hot gaseous emission is 
particularly luminous in the regions with the most active star formation
(Fabbiano et al.\ 2001; Zezas et al.\ 2002a, b; Zezas \& Fabbiano 2002; 
Fabbiano et al.\ 2003c). Based on these results, our team was awarded
a deep monitoring campaign of The Antennae with {\it Chandra}, which took
place in 2001--2002, and yielded a total 114~hr exposure, including
the original 1999 December data.
Preliminary reports of spectral and flux variability of the most
luminous X-ray sources in The Antennae are given in Fabbiano et al.\ (2003a, b).

In this Letter we present the entire deep data set
and give a first report on the spectral analysis of the diffuse emission,
made possible by the exceptional signal-to-noise ratio of our spectra.
The main result is the discovery of large spatial variations and enhancements
of the metal abundances.
Previous attempts at measuring the metallicity of the hot ISM with
the December 1999 data were
inconclusive (Fabbiano et al.\ 2003c).
We also report on the large-scale properties of the diffuse emission: 
While the first {\it Chandra} observations already suggested gaseous diffuse
emission south of the system (Fabbiano et al.\ 2001), the new deep
image reveals complex large-scale features.

\section{Deep {\it Chandra} Image of The Antennae}

Together with the original observation, the total {\it Chandra} ACIS-S
exposure on The Antennae is 411~ks, after background flare screening.
The standard Level 2 event files obtained by the Standard Data Processing 
(SDP) were used. The data set consists of seven observations 
($Obs ID$: 315, 3040, 3041, 3042, 3043, 3044, 3718). 
Figure~1 shows the astrometry-corrected and co-added `raw' data.
Details of the data reduction procedures are given in
Fabbiano et al.\ (2003a), Zezas et al.\ (2003, in preparation) and Baldi et al.\ (2003, in preparation). 
CIAO v3.0.1 and XSPEC were used for the data analysis. The latest ACIS gain
calibration was applied. Figure~2 shows an 
adaptively-smoothed true-color image, obtained from data  
in the 0.3--0.65, 0.65--1.5 and 1.5--6.0~keV energy bands.
This spectacular image shows both a population of point-like sources
and diffuse emission with a variety of spatial scales and spectral colors.
We will discuss elsewhere our detailed results on
the point-source population (Zezas et al.\ 2003, in preparation). 
In summary, we detect 120 sources ($3\sigma$), with 0.3--10.0~keV
luminosities in the range
$\sim 2 \times 10^{37}$\,--\,$5 \times 10^{39}$~\erg\, for 
an assumed power-law
spectrum with $\Gamma = 1.7$ and $N_{\rm H} = 3.4 \times 10^{20} \rm cm^{-2}$
(Galactic). Of the sources more luminous than the 
Eddington luminosity of a neutron star ($3 \times 10^{38}
\rm erg~s^{-1}$), $\sim$70\% are variable. 

Below, we concentrate on the most striking features of the diffuse soft
emission, which is detected mostly below 1.5~keV. Because of its spectrum,
this emission is likely to
be thermal emission from hot plasmas (Fabbiano et al.\ 2001, 2003c).
Figures~1 and 2 show that this emission is most intense in those areas
within the optical bodies of the galaxies that coincide with the most actively
star-forming regions. Figure~2 illustrates that this hot ISM is rich 
in features, both spatially and spectrally, as suggested by the varied 
range of X-ray colors. Larger-scale, more diffuse emission
is also present, and includes two 10~kpc-size 
(110\arcsec\ at D=19~Mpc) loop-like structures extending to the south of the main body
of emission.

\section{Metal Abundances of the Hot ISM}

The X-ray spectra extracted from different regions of the hot ISM
show a variety of continuum shapes and line-emission strengths (see Baldi
et al.\ 2003 for a complete study).  In particular, besides the Fe-L complex
($\sim$0.65--1.15~keV), 
lines of Ne X (0.92~keV), Mg XI-XII (1.34--1.47~keV), and Si XIII-XIV 
(1.85--2.01~keV) are visible in some of the spectra.
Figure~3 shows a {\it metallicity map} of the hot ISM.
The abundance of the above elements varies
widely in different regions, exceeding in places that of our local solar
neighborhood. 
{\it Chandra} observations of the dwarf starburst galaxy NGC~1569 have
led to the detection of large-scale metal-enriched hot winds (Martin et al.\
2002).  Our deep {\it Chandra} data allow for the first time the detailed
correlation of metal abundances with different galaxian regions and the
stellar populations.

Table~1 summarizes the
spectral properties of three  regions (out of 21
regions studied in Baldi et al.\  2003), representative of the range of
abundances. They are marked in
Figure~3 with `1', `5' and `7'.
We follow the numbering convention of Baldi et al.\ (2003)
for these regions, and we name them with R1, R5 and R7 in this paper. 
We fitted the spectra using the most 
complete thermal model available to us, the APEC model (Smith et al.\ 2001).
An intrinsic variable N$_H$ was fitted, in addition to the fixed line-of-sight
Galactic N$_H$. A power-law was also included in the fits to represent  harder sources
such as undetected X-ray binaries or young supernova remnants (SNR), and
yielded best-fit $\Gamma$ in the range 1.8--2.
Table~1 also lists mean stellar ages of the regions, estimated via a 
number-weighted average of the mean ages of highly reddened, extremely
young clusters (4 Myr; Zhang, Fall, \& Whitmore 2001)
and young blue clusters (15 Myr; Whitmore et al.\ 1999), and
the relative strength of the star formation rate (SFR) as measured by
the index  (U+H$\alpha$)/(V+I). 

R1 consists of two metal-poor regions in the northern loop of the
disk of NGC~4038. Figures~1 and 2 show a plume extending out of NGC~4038
near the western region, suggesting possible outflows.
R5, within the disk of NGC~4038, is a half-shell feature
of the hot ISM and has the most impressive emission lines. The spectrum and 
best-fit model (with residuals) are shown in Figure~4. 
This region is also the one with the highest relative SFR and, given its
small area, the largest SFR per unit area.
R7 is the absorbed region corresponding to the CO-rich ``Overlap
Region,'' where the most recent, intense star formation is occurring
(Wilson et al.\ 2000; see Fabbiano et al.\ 2003c).
The mean age of this region is based on optically detected clusters, so it may
represent an upper limit on the stellar age of the inner obscured regions. 
The relative SFR is large, but not as high as that of R5.

In both R5 and R7 the $\alpha$-elements are relatively more prominent than Iron, consistent with
enrichment by the SNe of type II expected in a young stellar population.
While we cannot draw any strong conclusions based on only three points,
Table~1 may suggest a trend between relative SFR and $\alpha$-element
abundances. 
The low abundances of R1 may also in part be due to the presence of
outflows, which
may disperse the element-enriched hot ISM outside of the galaxy.
 
While the presence of metals in the hot ISM is certain, the values of the
abundances are somewhat model-dependent.  For the 
static model, the radiative cooling time ($2 \times 10^7$ years) is longer than
the relevant recombination time ($2 \times 10^5$ years), and it is 
self-consistent to assume ionization equilibrium.  If the X-rays
come from a collection of SNRs, there 
could be some departure from equilibrium, but overionized
and underionized regions will tend to cancel out.  
If there is a wind, there
could be significant adiabatic cooling, so the ionization state
would be that appropriate for a higher temperature and abundances
could be overestimated (e.g., Breitschwerdt 2003).
That might affect the derived abundances for the low-temperature
regions far out in the halo, but high density regions such as R5 should not
be greatly affected.
R5 is remarkable for its high concentration of young star clusters.
It is possible that the
young clusters are still injecting kinetic energy predominantly by means
of stellar winds in this region, and that the temperature will increase
as the SN rate increases (Gilbert 2002).  It seems equally
likely
that the heated gas has entrained a large amount of cooler material, and
that as the region evolves the mean 
temperature will increase as the cool material is blown away.

\section{Large-scale Diffuse Emission}

At larger radii, Fig.~2 shows soft diffuse emission
without any obvious stellar/optical counterparts (Ponman et al.\ 2003, in preparation).
We can identify two components of this emission: (i) two spectacular extended structures,
suggesting giant loops or bubbles 
of hot plasma (the loops, hereafter), $\sim$10~kpc across, extending South of NGC~4039, and
(ii) a low-surface-brightness soft halo 
in the region surrounding the stellar disks of the merging galaxies, extending 
out to at least 18~kpc ($\sim$200\arcsec) from the nucleus of NGC~4039. The latter
is not clearly shown in Figure~2, but is revealed at high significance
by comparisons of the radial distribution
of the X-ray counts, with those expected from a field background (Fig.~5).

The spectra of the loops can be fitted with a thermal plasma with kT varying
between 0.29 keV and 0.34 keV ($1 \sigma$ errors of typically 0.01 to
0.03~keV) and metallicities apparently sub-solar (typically 0.1--0.2 solar).
The extent and spectral parameters of the low-surface-brightness halo are
uncertain because of the uncertainties of the background subtraction at these
large radii. However, the spectra are soft, with nominal
$kT=0.23+0.02/-0.01$~(1$\sigma$)~keV, so  
this larger-scale emission may be cooler than that of the loops.
The best-fit metallicity is only 0.04 solar, and is $<0.13$ at 90\% confidence
in the outer annulus. 
However, the signal-to-noise ratio of the data is lower than for the hot ISM,
and the fits include data from large regions.
While we have no reason to believe that the intrinsic spectral parameters
vary significantly at these radii, we cannot exclude it either.
Hence, these abundances are uncertain.
Taken at face value, there is a trend of decreasing abundances going from
the hot ISM of the star-forming regions to the loops, and to the larger-scale
emission. This trend may be consistent with a picture in which the
hot gas in the large-scale halo may  have been diluted with metal-poor
ambient ISM. The temperature also seems
to decrease at large radii, which may be due to adiabatic expansion.

What causes these large-scale features? The extended halo may be the
aftermath of superwinds from The Antennae, perhaps resulting from a
starburst at the first-encounter epoch $\sim$(2--5)$\times 10^8$yr ago
(Barnes 1988; Mihos et al.\ 1993; Whitmore et al.\ 1999) and the
prolonged elevated star-formation rate (4\,--\,6 \msun/yr) of the past
$\sim$150 Myr (Mihos, Bothun, \& Richstone 1993, esp.\ Fig.\ 10).
The cooling times in these diffuse regions are $\sim$1~Gyr. The thermal
energy content of both halo and loops is $\sim$10$^{56}$~erg,
comparable to the energy of $10^5$ SNe. Given the age of the
starburst of $\sim$10$^8$yr, a SN rate of 0.001 per year could
have provided this energy. This is a small fraction of the current
SN rate of The Antennae (0.2--0.3~yr$^{-1}$; Neff \& Ulvestad
2000) and, likely, of the average rate during the past $\sim$150 Myr.

A simple order-of-magnitude estimate of the expansion velocity of the loops,
assuming a 10~kpc extent and constant velocity over a period of
$10^7$\,--\,$10^8$ yr, yields a velocity of 1000--100~km/s.  At least
velocities $\ga$300~km/s are likely to exceed the escape velocity of
the system and correspond, in fact, to about the mean expansion velocity
observed in ultraluminous infrared galaxies (Rupke, Veilleux, \&
Sanders 2002).
Moreover, these velocities are comparable to or exceed 
the sound speed of $\sim$200~km~s$^{-1}$ we derive for this hot gas.
Therefore, the loops could be shocked swept-up ISM. 
If they are superbubbles
blown by a violent localized starburst (e.g., at the nucleus of NGC~4039),
as, e.g., in the Tomisaka \& Ikeuchi (1988) model, one would expect more of
a bipolar outflow morphology. They could correspond to two separate events,
with perhaps the receding
sides of the outflows obscured by the main body of the emission from the merger.
Another possibility that will need to be explored with hydrodynamical simulations,
is that the loops are somehow connected with the merging process, as
suggested in the case of Arp~220, where similar structures are observed
with {\it Chandra} (McDowell et al.\ 2003). We note that a
large-scale ring of hot gaseous emission is present in NGC~5128, a
likely merger remnant (Karovska et al.\ 2002). 

Given the long cooling times of both loops and halo in The Antennae, it seems
possible that---if these features are bound by a dark massive halo---they
may persist after the completion of the merger and the formation
of the resulting elliptical galaxy.  Interestingly, the  total mass of hot gas
in these extended features is $\sim${}$4 \times 10^8 M_{\odot}$.
This mass is comparable with that reported from X-ray observations of 
relatively X-ray faint Elliptical and S0 galaxies (e.g., Roberts et al.\ 1991).

\section{Conclusions}

The above results demonstrate the power of sensitive, high-resolution X-ray 
observations. Thanks to a deep 411~ks observation of The Antennae, we now have 
a first picture of the varying chemical enrichment
of the ISM, and we can see this hot medium expanding
into the intergalactic space and possibly enriching the intergalactic medium
with metals. Given the long cooling time of the extended halo, these observations
may also show the formation of the hot halo of the elliptical galaxy that will be 
created by the merger. These data provide a new, welcome opportunity to witness
the chemical evolution of galaxies in action, and offer a nearby glimpse
of the distant evolving Universe. 

\acknowledgments

 We thank the CXC DS and SDS teams for their efforts in reducing the data and 
developing the software used for the reduction (SDP) and analysis
(CIAO). We thank Martin Elvis and Larry David for useful discussions. 
This work was supported by NASA contract NAS~8--39073 (CXC)  and NASA
Grant G02-3135X.
ARK gratefully acknowledges a Royal Society Wolfson Research Merit Award,
and FS support from NSF grant AST-02\,05994.

{}

\clearpage
\begin{deluxetable}{lccccccccc}
\tabletypesize{\scriptsize}
\tablecaption{\small Best-fit (68\% errors) and Stellar Parameters for three Regions
of the
Antennae ISM\label{mytable}}
\tablewidth{0pt}
\tablehead{
\colhead{$Reg.^a$} &
\colhead{$\chi^2/dof$} &
\colhead{\begin{tabular}{c}
N$_H$\\
($10^{21}$cm$^{-2}$)
\end{tabular}} &
\colhead{\begin{tabular}{c}
$kT$\\
(keV)
\end{tabular}} &
\colhead{\begin{tabular}{c}
$Z_{Ne}$\\
($Z_{Ne,\odot}$)
\end{tabular}} &
\colhead{\begin{tabular}{c}
$Z_{Mg}$\\
($ Z_{Mg,\odot}$)
\end{tabular}} &
\colhead{\begin{tabular}{c}
$Z_{Si}$\\
($ Z_{Si,\odot}$)
\end{tabular}} &
\colhead{\begin{tabular}{c}
$Z_{Fe}$\\
($ Z_{Fe,\odot}$)
\end{tabular}} &
\colhead{\begin{tabular}{c}
Age\\
(Myr)
\end{tabular}} &
\colhead{\begin{tabular}{c}
Rel. \\
SFR\end{tabular}}
}
\startdata
R1$^b$&24.6/34&
$ 0.18_{-0.18}^{+0.73}$
&
\begin{tabular}{c}
$0.18_{-0.05}^{+0.04}$\\
$0.56_{-0.07}^{+0.05}$
\end{tabular} 
&$0.93_{-0.67}^{+2.02}$&$0.70_{-0.54}^{+0.93}$&$<0.25$&$0.58_{-0.27}^{+0.88}$ 
& 4 & 0.4\\
R5&31.5/34&$ 0.08_{-0.08}^{+0.23}$&$0.30\pm0.02$&$7.88_{-1.90}^{+1.43}$&$16.91_{-4.25}^{+7.64}$&$23.70_{
-10.73}^{+13.38}$&$2.67_{-0.34}^{+1.06}$ & 10.6 & 2.3\\
R7&83.2/73&$1.76_{-1.09}^{+0.70}$&$0.50_{-0.10}^{+0.08}$&$3.11_{-1.19}^{+4.15}$&$3.19_{-1.62}^{+4.97}$&
$4.40_{-2.41}^{+18.04}$&$0.78_{-0.41}^{+0.59}$ & 4 & 1.0\\
\enddata

a) R1 and R7 are representative low and high abundance regions out of 21 regions studied in 
detail by Baldi \etal\ (2003).\\
R5 is the region with the most extreme high abundances. These regions are identified by `1', `5' 
and `7' in fig.~3.

b) Two thermal components are required to fit the R1 data. While R1 consists of two areas, 
either one does not 
have \\
enough counts for a meaningful fit. Both areas have similar X-ray colors, suggesting similar spectra.

\end{deluxetable}

\setcounter{figure}{0}

\begin{figure}
\caption{Coadded astrometry-corrected `raw' data  (0.3-6 keV), together with a
representative outline of the optical interacting disks (blue contours),
and red contours derived from the {\it Hubble} WFPC-2 H$\alpha$ image
(from Whitmore et al.\ 1999), that indicates the position of HII 
regions in the Antennae. 
}
\end{figure}

\begin{figure}
\caption{Exposure-corrected 
adaptively-smoothed true-color image, obtained from data  
in the 0.3--0.65 (red), 0.65--1.5 (green) and 1.5--6.0~keV (blue) energy bands.
The data processing is similar to that followed in Fabbiano et al.\ (2003c) and is 
explained fully in Baldi et al.\ (2003, in preparation). The maximum significance 
has been set to 5$\sigma$, while the maximum smoothing 
scale corresponds to a minimum significance of 2.4$\sigma$ inside the optical 
body of the Antennae in the `green' band. The same smoothing scales (1 - 128 pixels)
were applied to the
three bands. The color scale
 has been stretched to allow the display of both the 
large-scale diffuse emission and the luminous, more resolved features.}
\end{figure}

\begin{figure}
\caption{Metallicity map of the hot ISM of
The Antennae, where  red, green, and blue  indicate emission by
Fe, Si, and Mg respectively.
Continuum was subtracted by estimating the contribution of the overall
best-fit two-component thermal bremsstrahlung model in the two  bands 
 at 1.4--1.65~keV and 2.05--3.50~keV, where no strong lines are observed.
 Point-like sources were excluded, following Fabbiano et al.\ (2003c).
The Fe-L band image was
adaptively smoothed  to a significance between 3$\sigma$ and
5$\sigma$, and the same scales were applied to the other line images.}
\end{figure}

\begin{figure}
\caption{ACIS spectrum of region R5, together with best-fit model and fit residuals.
Gain corrections were applied using software provided by the CXC Calibration Group.}
\end{figure}

\begin{figure}
\caption{Surface brightness (0.3 - 1.0 keV, solid line), compared with that of the expexted 
field background (dashed)}
\end{figure}


\begin{thebibliography}{}

\bibitem[]{} Baldi, A. et al.\ 2003, in preparation

\bibitem[Barnes(1988)]{barn88} Barnes, J.\ E. 1988, \apj, 331, 699

\bibitem[]{}Breitschwerdt, D. 2003, RMxAC, 15, 311

\bibitem[Fabbiano et al.\ 2001]{}Fabbiano, G., Zezas, A., \& Murray, S.
2001, \apj, 554, 1035 

\bibitem[Fabbiano \etal\ 2003]{}Fabbiano, G., Zezas, A., King, A. R.,
Ponman, T. J., Rots, A., Raymond, J., \& Schweizer, F. 2003a,
\apj, 584, L5

\bibitem[]{} Fabbiano, G., King, A. R., Zezas, A., Ponman, T. J., Rots, A., Schweizer, F. 2003b,
\apj, 591, 843

\bibitem[]{} Fabbiano, G., Krauss, M., Zezas, A., Rots, A., \& Neff, S. 2003c, \apj, in press

\bibitem[]{} Fabbiano, G. \& Trinchieri, G. 1983, \apj, 266, L5

\bibitem[]{}
Gilbert, A.M., 2002, PhD Thesis, University of California, Berkely

\bibitem[Hibbard et al 2001]{876} Hibbard, J. E., van der Hulst, J. M.,
Barnes, J. E., \& Rich, R. M. 2001, AJ, 122, 2969

\bibitem[Karovska et al 2002]{879} Karovska, M., Fabbiano, G., 
Nicastro, F., Elvis, M., Kraft, R. P. \& Murray, S. S.
2002, \apj, 577, 114

\bibitem[]{}Martin, C. L., Kobulnicky, H. A. \& Heckman, T. M. 2002, \apj, 574, 663

\bibitem[]{}McDowell, J. C., et al. 2003, \apj, 591, 154

\bibitem[Mihos et al.(1993)]{miho93} Mihos, J.\ C., Bothun, G.\ D., \&
   Richstone, D.\ O. 1993, \apj, 418, 82


\bibitem[]{}Navarro, J. F., Frenk, C. S., \& White, S. D. 1995, \mnras, 275, 56

\bibitem[Neff \& Ulvestad(2000)]{neff00} Neff, S.\ G., \& Ulvestad, J.\ S.
   2000, \aj, 120, 670


\bibitem[]{} Ponman, T. et al.\ 2003, in preparation

\bibitem[]{} Roberts M. S., Hogg, D. E., Bregman, J. N., Forman, W. R., Jones, C. 1991, \apjs, 75, 751


\bibitem[]{} Rupke, D.\ S., Veilleux, S., \& Sanders, D.\ B. 2002, \apj, 570, 588


\bibitem[]{}Smith, R. K., 
 Brickhouse, N. S.,
 Liedahl, D. A., Raymond, J. C. 2001, \apj, 556, L91

\bibitem[]{} Tomisaka, K. \& Ikeuchi, S. 1988, \apj, 330, 695



\bibitem[Weisskopf \etal\ 2000]{939}Weisskopf, M., Tananbaum, H., Van Speybroeck, L. \& O'Dell, S.
2000, Proc. SPIE 4012, p. 2 (astro-ph 0004127)

\bibitem[Whitmore et al.(1999)]{whit99} Whitmore, B.\ C., Zhang, Q.,
   Leitherer, C., Fall, S.\ M., Schweizer, F., \& Miller, B.\ W. 1999,
   \aj, 118, 1551


\bibitem[Wilson et al 2000]{952}Wilson, C. D., Scoville, N., Madden, S. C., \& 
Charmandaris, V. 2000, \apj, 542, 120


\bibitem[Zezas \& Fabbiano 2002]{} Zezas, A. \& Fabbiano 2002, \apj, 577, 726

\bibitem[Zezas \etal\ 2001]{} Zezas, A., Fabbiano, G. Rots, A. H., \&
Murray, S., 2002a, \apjs, 142, 239

\bibitem[Zezas \etal\ 2001]{} Zezas, A., Fabbiano, G. Rots, A. H., \&
Murray, S., 2002b, \apj, 577, 710

\bibitem[]{}Zezas, A. et al.\ 2003, in preparation

\bibitem[Zhang et al 2001]{968}Zhang, Q., Fall, S. M., \& Whitmore, B. C.
2001, \apj, 561, 727

\end{thebibliography}
\end{document}